# Geometrical and orbital effects in a quasi-one dimensional conductor

D. Graf,[1] J.S. Brooks,[1] E. S. Choi,[1] M. Almeida,[2] R.T. Henriques,[2] J.C. Dias,[2] and S. Uji[3]

[1]Department of Physics/Chemistry and National High Magnetic Field Laboratory,
Florida State University, Tallahassee, FL 32310, USA
[2]Instituto Tecnológico e Nuclear / CFMCUL, Estrada Nacional n 10, P-2686-953 Sacavém, Portugal
[3]National Institute of Materials Science, Tsukuba, Ibaraki, 305-0003 Japan

The angular dependent magnetoresistance of (Per)$_2$[Au(mnt)$_2$] under pressure has revealed geometrical effects associated with the crystallographic parameters. Pressure suppresses the charge density wave ground state of the material, and in the metallic state both geometrical and orbital quantum interference effects appear. Through magnetic field dependence and orientation, the orbital and geometrical effects are independently identified. We compare the results from (Per)$_2$[Au(mnt)$_2$] with previous studies of the well-known Bechgaard salts.

## I. Introduction

Low dimensional organic conducting materials exhibit a set of magnetotransport properties associated with their quasi-one and two dimensional Fermi surface (FS) topologies. Fermiology generally describes the quantum oscillation behavior associated with closed orbits where de Haas - van Alphen related measurements yield the area of extremal orbits in k-space, and angular dependent magnetoresistance (ADMR) effects yield additional details of the orbital topology. In the case of quasi-one-dimensional (Q1D) organic metals, the open Fermi surfaces are generally warped by the smaller transverse bandwidths of the material that modify, but are too small to close the Fermi surface in those directions. For carrier trajectories along Q1D FS sheets different classes of ADMR phenomena exists, commonly called Lebed magic angle[1,2], Danner-Kang-Chaikin[3], and third angle[4] effects. These Q1D effects are now well established in the literature for both the Bechgaard salts[5,6] and also to a large extent in the DMET salts[7], both of which share a similar electronic anisotropy. In the case of Lebed magic angles, magnetoresistance anomalies occur when the field is aligned in a direction involving an integer ratio of the transverse unit cell parameters. This effect was first proposed by Lebed, *et al.,* as a resonance of the motion of carriers across the Brillouin zones of Fermi surface sheets. The origins of these angular effects have remained controversial[6,8] and open to interpretation.

The work presented here explores ADMR effects in the case of (Per)$_2$[Au(mnt)$_2$]. Unlike the Bechgaard salt (TMTSF)$_2$ClO$_4$ where the metallic state exhibits magic angle effects below a threshold field for field-induced spin density wave formation, the charge density wave (CDW) in (Per)$_2$[Au(mnt)$_2$] must first be suppressed either by high magnetic fields[9] or pressure[10,11]. Under increasing pressure, manifestations of the underlying metallic quasi-one-dimensional electronic structure as shown in Fig. 1a, emerge. Indeed, for a pressure in the range of 5 kbar, (Per)$_2$[Au(mnt)$_2$] is fully metallic from room temperature to 0.3 K, below which superconductivity appears[12]. The interchain transfer energies associated with the least conducting c and a directions are estimated as 0.1 and 2.0 meV respectively at ambient pressure[13], i.e. the warping of the FS sheets is about an order of magnitude lower that in the Bechgaard salts[14]. Due to the non-equivalent perylene interchain interactions within each unit cell, there are as many as four open orbit sheets near $\pm k_F$[13] quite similar to the corresponding pair of FS sheets at $\pm k_F$ in (TMTSF)$_2$ClO$_4$ that arise due to the doubling of the unit cell from anion ordering. In both materials, this topology leads to quantum interference (QI) orbits in the metallic state[11,15].

## II. Experimental Details

Single crystal samples of (Per)$_2$[Au(mnt)$_2$] with typical dimensions of 2 x 0.05 x 0.02 mm$^3$ (corresponding to the b, a, and c-axes respectively) were grown electrochemically[16]. Electrical contacts were made using 12 μm Au wire and carbon paint. Typically, several samples with different orientations were mounted in a BeCu double-clamped pressure cell using Daphne 7373 oil as a pressure medium. The pressure clamps were affixed to rotatable stages in helium-3 cryostats in both 18 T superconducting and 31 T resistive magnets. The inset to Fig. 1c defines the direction of magnetic field rotation with respect to the crystal axes. Low frequency ac resistance measurements were made with a standard 4-terminal lock-in configuration for current (I ~ 10 µA) directed in either the ac-plane, or along the b-axis.

### III. Results and Discussion

The main challenge in identifying the geometrical effects associated with the FS sheet topology in $(Per)_2[Au(mnt)_2]$ is to distinguish the ADMR signatures from those arising from QI orbits[11]. In Fig. 1b we show the ADMR for rotation between the ac-plane and the b-axis for several constant magnetic fields, where pronounced QI effects are observed. In the inset to Fig. 1b, data between $90^o$ and $180^o$ are displayed vs. $B\cos(\phi)$ and compared with data taken vs. field for $B \perp b$. It is clear from Fig. 1a that the effects of rotation from the b-axis to the ac-plane will follow a $B\cos(\phi)$ dependence since the effective extremal area encompassed by the QI orbits increases when rotating towards the ac-plane, regardless of the ac-plane field projection. In Fig. 1c, a dramatically different, complex behavior is observed in the ADMR for a θ rotation in the ac-plane. First, there appear a series of dip-features whose angular positions are independent of field (geometrical effects, hereafter labeled 1 to 6); second there are less pronounced dip-features that are dependent on field (orbital effect); and third the background resistance (MR effect) also change with field. The reproducibility of complex behavior is demonstrated by comparing the two traces at 18 T in Fig. 1c. Here two samples measured in the same pressure cell, but with different ac-plane orientations (offset $68^o$) and different current directions are compared. By accounting for the offset, the ADMR signals from the independent samples show a nearly identical θ dependence. A more systematic ADMR study is shown in Fig. 2a where both the geometrical and orbital features can be followed in more detail with magnetic field.

The Lebed-type geometrical features depend only on the field direction through $\tan(\theta) = p c \sin(\beta)/(qa + pc \cos(\beta))$ where $p$ and $q$ are integers, and $\beta$ is the angle between the a and c lattice parameters. In contrast, ADMR features associated with the quantization of orbits will depend on the component of the field perpendicular to the effective extremal orbital area. In the case of QI orbits that require only multiple FS sheets with similar Fermi momenta, as in Fig. 1a, there is no orientation where the oscillations associated with the QI orbits disappear with angle θ; rather, their quantization condition will depend on the effective QI extremal area determined by the warped FS topology, which will alternate between *B // a* and *B // c*. *The following point is therefore of central importance*: if a particular ADMR feature is followed with increasing field, and if the angular (θ) position changes with field, then it is an extremal area effect associated with the QI oscillations. However, if the position in θ of an ADMR feature is field-independent, then the feature is associated only with the lattice parameters, i.e. a geometrical effect.

In light of the above, the behavior of both the geometrical and orbital effects can be summarized as a polar plot shown in Fig. 2b, where the main magnetic field dependent (orbital) and magnetic field independent (geometrical) features vs. θ derived from Figs. 1c and 2a are presented. Turning first to the orbital features, we find that their positions with field and angle follow roughly two symmetrical sets, one centered in the effective a-direction, and the other centered around the c-direction. We assert that these features are associated with quantization of the extremal QI orbits. Considering the main dip in the MR at $B_\perp = 9.5$ T from the inset of Fig. 1b, we may write the condition for quantization as $B = |B_\perp/\cos(\theta - \theta_0)|$, which is presented in Fig. 2b as the parallel dashed lines. ($\theta_0$ relates the experimental and crystallographic angles.) This orbital feature corresponds to the N = 2 dip in the QI signal[11]. For values of B and θ that satisfy the quantization condition, there will be a constructive QI component $B_\perp$ normal to the a-direction. This relationship is shown by schematic representation of the effective extremal orbits associated with the $A_{ab}$ and $A_{bc}$ k-space areas in the inset of Fig. 2b; hence a similar relationship will occur for the c-direction. Note that at low fields, the quantization condition cannot be met and the orbital features disappear.

We next consider the field independent geometrical features (numbered 1 to 6), and have correlated their field directions with the $(Per)_2[Au(mnt)_2]$ ac-plane crystal structure as shown in Fig. 3. The notation and values for the transfer integrals have been taken from Ref. [13]. Many of the magnetoresistance features can be explained directly from the geometry of the crystal structure. For example, when the field is well-aligned with the transfer integrals (dips 1, 3 and 5), we find the most pronounced effect in high fields, noting that the most distinct dip occurs with alignment along the strongest transfer integrals (5: a-axis). When the field is aligned with a path that is less direct (dips 2, 4 and 6) the features are less dramatic. We may also apply the Lebed relationship above to correlate the MA features with the integer lattice constant ratios. Following the convention *[q(a-axis), p(c-axis)]*, we find that *[0a,c], [1a,-1c], [2a,-1c], [4a,-1c], [1a,0c], and [2a,1c]* correspond to the directions *1 (c-axis), 2, 3, 4, 5(a-axis), and 6* respectively.

The overall background ADMR of (Per)$_2$[Au(mnt)$_2$] can also be compared with that of (TMTSF)$_2$PF$_6$. First, (TMTSF)$_2$PF$_6$ has nearly orthogonal symmetry in the bc-plane (dips for directions b-c-b appear at -90$^o$, 8$^o$, +90$^o$) whereas the honeycomb-like structure of the perylene donors of the ac-plane of (Per)$_2$[Au(mnt)$_2$]creates the main dips for a-c-a at -60$^o$, 0$^o$, +120$^o$ with higher order commensurate dips spaced in between at roughly 30$^o$ intervals. Therefore a comparison of the *B* - θ nature of the background MR treated by Strong *et al.*[17] is more difficult to make at this point.

The temperature dependence of the magic angle dips is also of importance. In the case of (TMTSF)$_2$PF$_6$, it was shown that the dips have a metallic (*dR/dT* > 0) behavior, whereas away from the dips *dR/dT* < 0, due to coherently coupled or decoupled layers, respectively[18]. The temperature dependence of the ADMR for the B//a-axis (#5) orientation is shown in Fig. 4. The resistance at the dip ($R_{dip}$) is assumed to behave as a parallel combination of the more resistive part ($R_{max}$) and a metallic contribution ($R_{metal}$). For the many samples we have measured at similar pressures, $R_{dip}$ is metallic or weakly activated, but it is consistently far less activated than for angles away from MA. As mentioned above (and also considered in Refs. 10,11), residual CDW resistive contributions in parallel or series and/or new Q1D field induced instabilities may contribute to a background magnetoresistance regardless of field direction. One final note concerns the dip at *B // b* (see θ = 90$^o$ in Fig. 1b). For this direction, the temperature dependence is strictly metallic and no detectable MR is observed.

A number of models have been developed to explain the magic angle effect in the Bechgaard salts[1,17-21] and a comparison with these materials provides some insight into the ADMR behavior of (Per)$_2$[Au(mnt)$_2$]. It has been suggested[17,22] the magic angle features at constant angles can be explained as follows: when the magnetic field is aligned in the bc-plane of (TMTSF)$_2$PF$_6$ the carrier movement becomes increasingly restricted along the c-axis as the field increases. The carriers are confined once the width of the real space trajectory is smaller than the width of the layer, described as a two dimensional (2D) crossover. The coherence of the carriers is decreased at orientations away from *B // c* or away from magic angles and the scattering is increased (the MR is larger). When the field is aligned with the magic angle orientations, carrier coherence is maintained and ADMR dips are observed. The confinement condition[18,22] can be estimated using the relation ev$_F$H$_{a,b}$c > $t_c$ where e is the charge, $v_F$ is the Fermi velocity, c is the real space width of the carrier trajectory and $H_{a,b}$ is the magnitude of the field aligned with the a- (or b-) axis of the perylene (or TMTSF) system. Both systems have Fermi velocities[8,13] of ~ 1.7 x 10$^5$ *m/s* and comparing the crystal parameters for both materials[14] indicates that confinement would occur in the perylene compound at magnetic fields significantly less ($H_{a,b}$ < 1*T*) than in (TMTSF)$_2$PF$_6$. It is therefore possible that the MA mechanism for (Per)$_2$[Au(mnt)$_2$] is somewhat different from that discussed above, especially since $t_c$ is already so small, i.e, the crossover is from 2D to 1D, not 3D to 2D. This could also lead to contributions in the temperature dependent background, if for instance, the field induces some kind of instability in the 1D system[23,24].

In summary, we have identified two angular dependent magnetoresistance effects in (Per)$_2$[Au(mnt)$_2$] under pressure. Orbital, quantum interference effects are observed for all field directions that produce extremal cross-sections through the Fermi surface (Fig. 1a). In contrast, the appearance of geometrical effects depends only on the orientation between the magnetic field (and not magnitude) and the crystal structure within the ac-plane. This discovery is an example of geometrical effects beyond the TMTSF based systems, and is the first such system where MA effects to be observed in a Q1D material with a CDW ground state. In a comparison between the Bechgaard and perylene salts, we note that there may be fundamental differences in the origin of one or more of the MA effects due to the extremely low transverse bandwidths and lower symmetry in the perylene salts.

Acknowledgments

This work was supported in part by NSF DMR-0602859 (JSB) and performed at the National High Magnetic Field Laboratory, supported by NSF DMR-0654118, by the State of Florida, and the DOE.

Figures

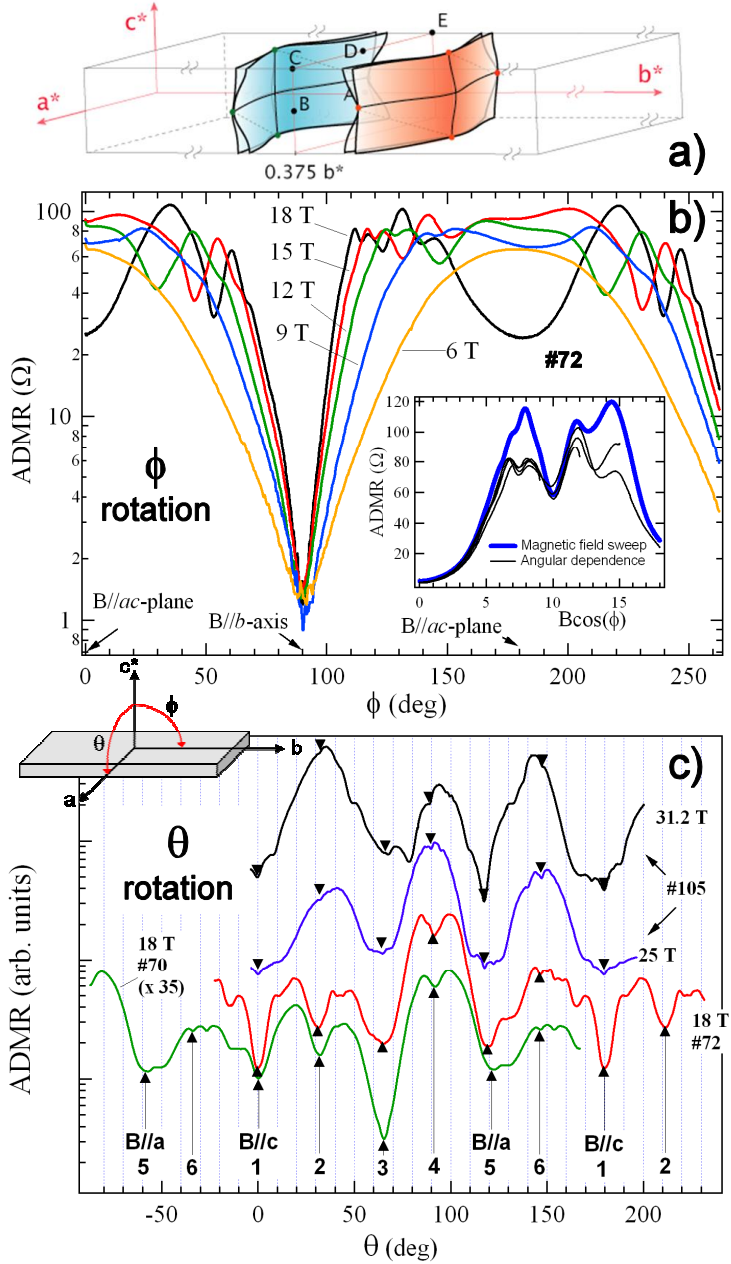

FIG. 1: (color online) (a) Fermi surface of (Per)$_2$[Au(mnt)$_2$] (redrawn after Ref. [13] - warping is exaggerated). Middle inset: Definition of angles $\phi$ and $\theta$ for field rotations between the b-axis and the ac-plane, and the a-axis and the c-axis, respectively. (b) ADMR of (Per)$_2$[Au(mnt)$_2$] (sample 72) at constant magnetic fields for $\phi$ rotations. (I⊥b, pressure = 6.4 kbar and T = 300 mK.) Inset: ADMR (thinner curves) plotted with respect to Bcos($\phi$) compared with magnetoresistance (thicker curve) for constant angle $\phi \sim 0^o$ (i.e. perpendicular to the b-axis). (c) ADMR of (Per)$_2$[Au(mnt)$_2$] at constant magnetic fields for $\theta$ rotations in the ac-plane for T ~ 300 mK . Pressure = 6.4 kbar (18 T data for samples 70 and 72) and 5.4 kbar (25T and 31.2T data for sample 105). I ⊥ b for samples 105 and 72; I ∥ b for sample 70 (trace magnified 35 times due to the smaller b-axis resistance). Log-scaled traces are offset for clarity. The numbered arrows refer to angular positions of the field-independent ADMR dip features.

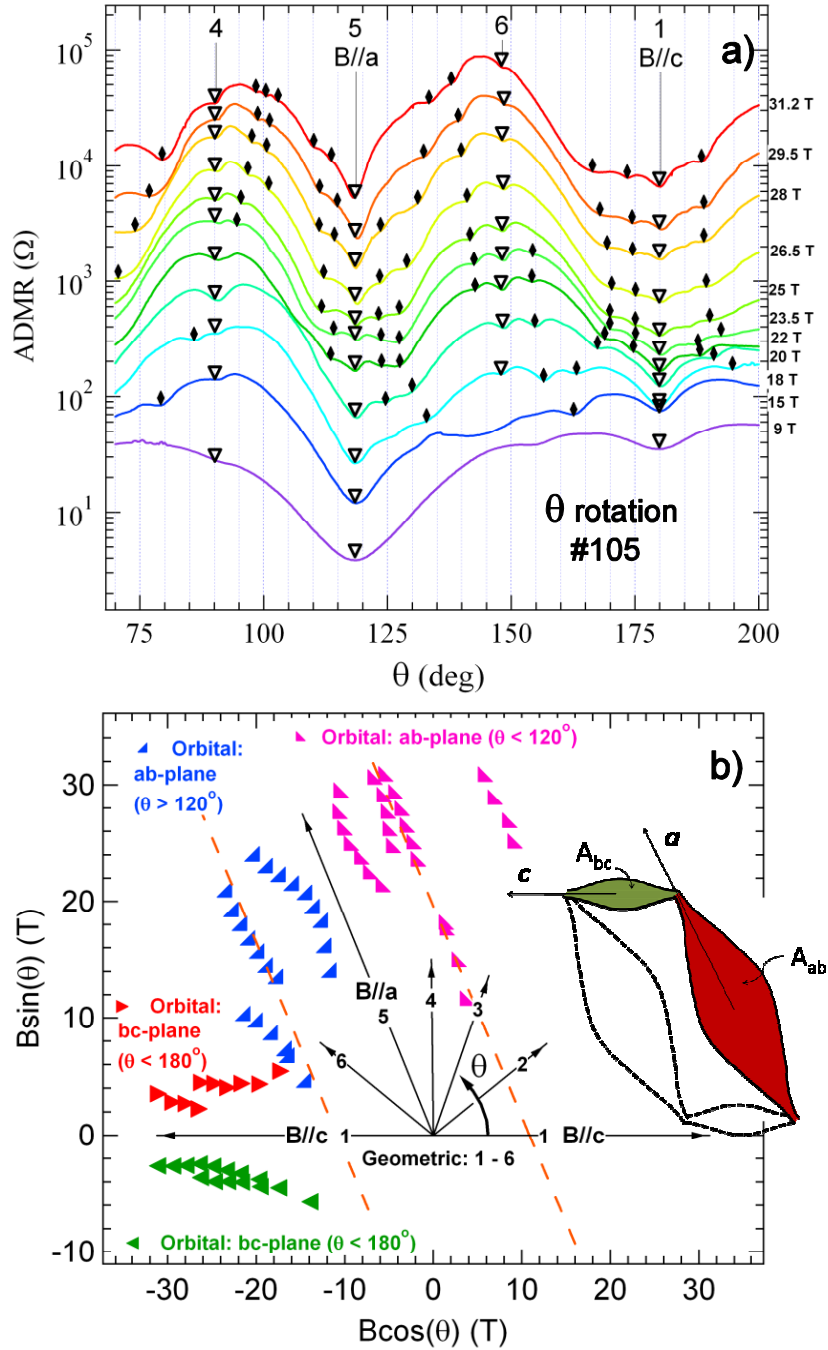

FIG. 2: (color online) (a) Systematic field dependent θ rotation study of $(Per)_2[Au(mnt)_2]$ sample 105 (5.4 kbar, 0.3 K). Larger, open numbered triangles indicate the location of the field-independent ADMR features; diamonds indicate the location of the field-dependent ADMR features. (b) Comparison of geometrical (field-independent, black arrow) and orbital (field-dependent, symbol) features vs. θ obtained from Fig. 1b and (a) above (note direction of increasing θ). Dashed lines: fit to the orbital QI feature associated with the 9.5 T dip feature (see text for discussion). The orbital features cluster around two different directions, indicating that the orientation of the effective QI extremal Fermi surface areas $A_{ab}$ and $A_{bc}$ are parallel to either the a- or c-axes respectively, represented schematically by the inset.

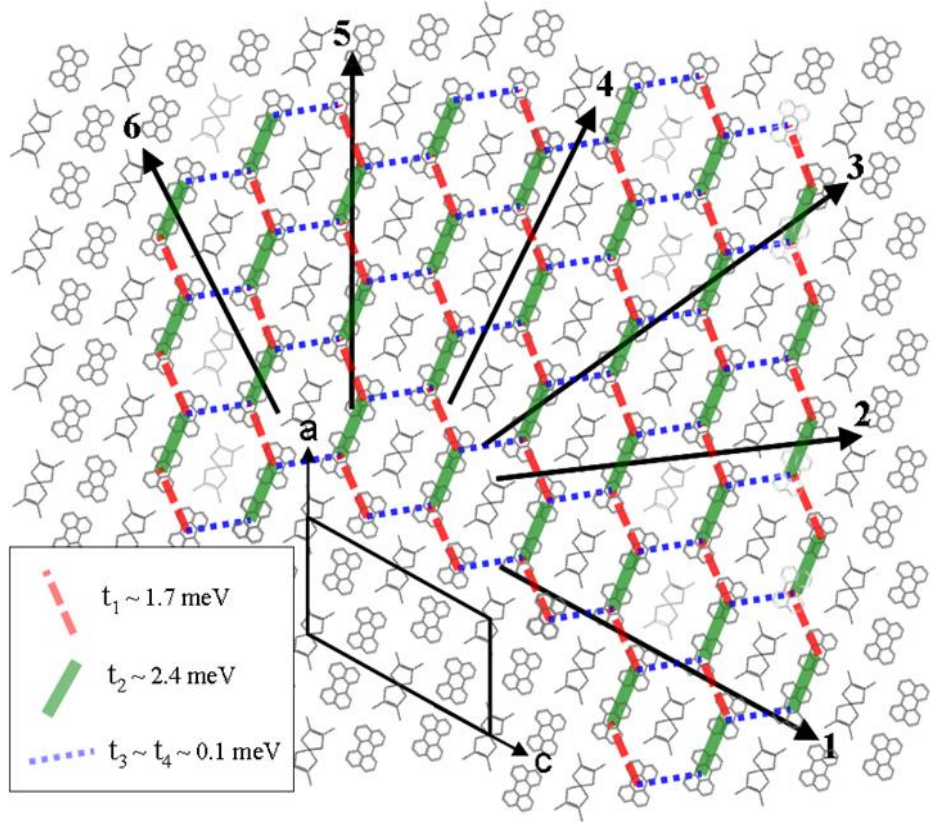

FIG. 3: (color online) Correlation of the geometrical (field-independent) features with the ac-plane crystal structure and transfer integral energies (Ref. [13]) of $(Per)_2[Au(mnt)_2]$. The ADMR dip features 1 - 6 in Figs. 1c and 2 correspond to alignment between the magnetic field and molecular planes defined by the black arrows. The thickness of the lines use for the transfer integrals indicates their relative magnitudes.

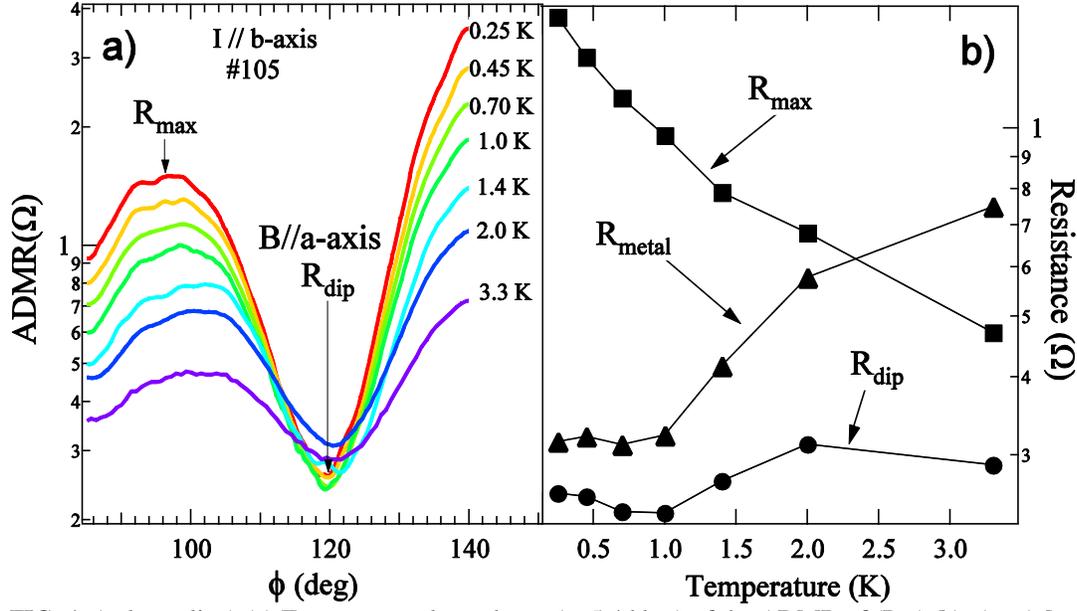

FIG. 4: (color online) (a) Temperature dependence (at 5.4 kbar) of the ADMR of $(Per)_2[Au(mnt)_2]$ near $B // a$ (crystallographic plane 5 in Fig. 3). (b) Comparison of the temperature dependence of the background resistance at $R_{max}$ (at a field position away from plane 5) and the resistance minimum at $R_{dip}$ for $B // a$. Here $R_{metal} = R_{max}R_{dip}=(R_{max} - R_{dip})$ is the effective metallic resistance, assuming parallel contrib-tions from the geometrical and background conductivities at the dip.